\documentclass[11pt]{article}
\usepackage{amssymb,cite}
\usepackage[dvips]{graphicx}
\makeatletter
\@addtoreset{equation}{section}
\makeatother

\let\bm=\bibitem

\title{\Large{Charges of Exceptionally Twisted Branes}}

\author{Muraari Vasudevan}

\newcommand{\auth}{\textbf{Terry Gannon$^*$} and \textbf{Muraari Vasudevan$^\dagger$}}

\begin{document}

\begin{flushright}

Alberta Thy 05-05\\
May\  2005
\end{flushright}

\vspace{10pt}

\begin{center}
{\Large {\bf Charges of Exceptionally Twisted Branes
            }}

\vspace{20pt}
\auth

\vspace{10pt}

{\it $^*$ Department of Mathematical Sciences,\\ University of Alberta,
Edmonton, Alberta  T6G 2G1, Canada \\
 {\rm E-mail: \texttt{tgannon@math.ualberta.ca}}
  }
\vspace{10pt}

{\it $^\dagger$ Theoretical Physics Institute,\\ University of Alberta,
Edmonton, Alberta  T6G 2J1, Canada \\
 {\rm E-mail: \texttt{mvasudev@phys.ualberta.ca}}
  }

\vspace{40pt}

\underline{ABSTRACT}
\end{center}

The charges of the exceptionally twisted (D4 with triality and E6
with charge conjugation) D-branes of WZW models are determined
from the microscopic/CFT point of view. The branes are labelled by
twisted representations of the affine algebra, and their charge is
determined to be the ground state multiplicity of the twisted
representation. It is explicitly shown using Lie theory that the
charge groups of these twisted branes are the same as those of the
untwisted ones, confirming the macroscopic K-theoretic
calculation. A key ingredient in our proof is that, surprisingly,
the G2 and F4 Weyl dimensions see the simple currents of A2 and
D4, respectively.

\pagebreak

\section{Introduction}

Conserved charges of D-branes in String Theory, to a very large part, determine
their effective dynamics. As such, determining these charges and the associated
charge groups provides significant information regarding the D-branes. For
strings propagating on a group manifold, i.e. a $g_k$-WZW model, these charges
can be determined using the underlying CFT \cite{GW}. WZW models possess an
extremely rich variety of D-brane dynamics directly attributable to the
additional affine Lie structure, which is preserved by the D-branes.

In addition to the standard untwisted branes, WZW models also
possess D-branes which preserve the affine symmetry only up to a
twist, the so-called ``twisted" branes. For every automorphism
$\omega$ of the finite dimensional Lie algebra $\overline{g}$ of the affine Lie algebra
$g$, there exist $\omega$-twisted D-branes. It is sufficient to
consider outer automorphisms only, and as such, only automorphisms
determined by symmetries of the Dynkin diagram of $\overline{g}$
\cite{FC,Kac}. Such twists exist for the $A_n$'s, $D_n$'s, and
$E_6$, where $\omega$ in each case is an order two symmetry referred
to as charge conjugation (or chirality flip in the case of $D_n$ with $n$ even), and for $D_4$, where
$\omega$ is an order three symmetry referred to as triality. The microscopic analysis for twisted D-branes started with \cite{ASpre}, and a study at large
affine level was done in \cite{AFQS}. The
charges and charge groups for the order-two twisted $A_n$ and
$D_n$ D-branes have been calculated in \cite{GG2} (up to some
conjectures). This paper deals with the remaining cases of $D_4$
with triality and $E_6$ with charge conjugation.

The computations for $D_4$ and $E_6$ presented here, are purely Lie
theoretic, and are done from a ``microscopic"/CFT point of view.
These calculations provide confirmations for the results for the
charge group obtained ``macroscopically"/geometrically using
K-theory \cite{VB}. However, the K-theoretic arguments only
determine the charge group and not the charges themselves, so the
calculations done here provide significantly more information about
the D-branes.

We also prove some %a number of remarkable
Lie theoretic identities % regarding affine Lie algebras,
which warrant further study. The most surprising, and likely important, of
these are that $G_2$ and $F_4$ see the simple currents of $A_2$ and $D_4$,
respectively. More precisely, for arbitrary choice of level $k$, the simple
currents $J^i$ of $A_2$ permute
the integral weights $a'$ of $G_2$ in such a way that
\begin{eqnarray}
\mathrm{dim}_{G_2} (J^i a')=\mathrm{dim}_{G_2} % (J^2 b')=\mathrm{dim}_{G_2}
(a') \quad \mathrm{mod}\, M_{G_2} \, ,
\end{eqnarray}
where %$J$ is the simple current of $A_2$,
$M_{G_2}$ is an integer given next section.
%characteristic of $G_2$ (rather than $A_2$), and
Similarly, the 4 simple currents $J$ of $D_4$ permute the integral weights
of $F_4$ in such a way that
\begin{eqnarray}
\mathrm{dim}_{F_4} (Jb')= \mathrm{dim}_{F_4} (b') \quad \mathrm{mod} \, M_{F_4} \, ,
\end{eqnarray}
where likewise $M_{F_4}$ is given next section.
%the $J's$ are any of the 4 simple currents of $D_4$, $M_{F_4}$ is an integer characteristic of $F_4$ (rather than $D_4$). %%

We first provide a brief summary of the description of untwisted
D-brane charges in CFT, as well as the order-two twisted
D-branes of $A_n$ and $D_n$. Subsequently, we deal with the
exceptional cases of $D_4$ and $E_6$. The %A number of
non-trivial Lie
theoretic identities, which are needed along the way, are stated and
proved in the appendices.

\section{Overview of WZW D-Brane Charges in CFT}

The WZW models of relevance here are the ones on simply connected
compact group manifolds (partition function given by charge
conjugation). D-branes that preserve the full affine symmetry are
labelled by the level $k$ integrable highest weight
representations $P_+ ^k (g)$ of the affine algebra $g$. They are
solutions of the ``gluing" condition
\begin{equation}
J(z)= \overline{J} (\overline{z}) \, , z=\overline{z} \, ,
\end{equation}
where $J, \overline{J}$ are the chiral currents of the WZW model
\cite{AS}.

The charge $\mathrm{q}_{\mu}$ of the D-brane labelled by $\mu$
satisfies
\begin{eqnarray}
\mathrm{dim} (\lambda)\, \mathrm{q}_{\mu} = \sum _ {\nu \in P^k _+ (g)}
N_{\lambda \mu} ^{\nu} \mathrm{q}_{\nu}\, \quad \mathrm{mod}\,  M\,
, \label{unteqn}
\end{eqnarray}
where $\lambda \in P^k _+ (g)$, $N_{\lambda \mu} ^{\nu}$ are the
$g_k$-affine fusion rules, and $\mathrm{dim}(\lambda)=\mathrm{dim}(\overline{\lambda})$ denotes the dimension of the $\overline{g}$
representation whose highest weight is the finite part of the affine weight
$\lambda$ --- in this paper we freely interchange the affine weight $\lambda$
with its finite part $\overline{\lambda}$, which is unambiguous since
the level will always be understood. For a finite level $k$, this relationship
 (\ref{unteqn}) %
is only true modulo some integer $M$, and the charge group of these
D-branes is then $\mathbb{Z}/M\mathbb{Z}$, where $M$ is the largest positive integer
such that (\ref{unteqn}) holds. We are assuming here that the
only common divisor of all the $\mathrm{q}_{\mu}$ is 1 (if they do have a common
divisor, then this factor can be divided out). Without loss of
generality, we assume the normalization $\mathrm{q}_0=1$. If we take $\mu$ to be the trivial representation 0, %
then clearly
\begin{eqnarray}
\mathrm{q}_{\lambda} = \mathrm{dim}({\lambda})\, . \label{untchar}
\end{eqnarray}
The integer $M$ is then the largest integer such that
\begin{eqnarray}
\mathrm{dim} (\lambda)\, \mathrm{dim} (\mu) = \sum _ {\nu \in P^k
_+ (g)} N_{\lambda \mu} ^{\nu} \mathrm{dim}({\nu}) \qquad \mathrm{mod} \, M\,%,
\end{eqnarray}
holds. It has been conjectured (and proved for the $A_n$ and the $C_n$ series) in \cite{FS,BDR,VB} that the integer $M$ is always of
the form
\begin{eqnarray}
M=\frac{k+h^{\vee}}{\mathrm{gcd}(k+h^{\vee},L)} \, ,
\end{eqnarray}
where $h ^{\vee}$ is the dual Coxeter number of $\overline g$ and $L$ is
%an integer for the various algebras (independent of the level $k$)
a $k$-independent integer given in Table \ref{table1}\footnote{It
was suggested initially in \cite{BDR} that there were exceptional
values for $M$ at low levels $k$. However, this issue was
subsequently resolved in \cite{VB}, where it was proved that there
are no exceptional cases using K-theory. We will show that there
are no exceptional cases at low levels in both algebras, on the
CFT and Lie theory side,  when we prove uniqueness of the
solutions. }.

%\vspace{3mm}
\begin{table}[h]
\begin{center}
\begin{tabular}{|c|c|c|c|}
\hline
$\mathrm{Algebra}$ & $h^{\vee}$ & $L$ \\
\hline
$A_n$ & $n+1$  & $\mathrm{lcm} (1,2,\ldots,n)$    \\
$B_n$ & $2n-1$ & $\mathrm{lcm} (1,2,\ldots, 2n-1)$\\
$C_n$ & $n+1$  & $2^{-1}\mathrm{lcm} (1,2,\ldots,2n)$ \\ %
$D_n$ & $2n-2$ & $\mathrm{lcm} (1,2,\ldots,2n-3)$ \\
$E_6$ & $12$   & $\mathrm{lcm} (1,2,\ldots,11)$   \\
$E_7$ & $18$   & $\mathrm{lcm} (1,2,\ldots,17)$   \\
$E_8$ & $30$   & $\mathrm{lcm} (1,2,\ldots,29)$   \\
$F_4$ & $9$    & $\mathrm{lcm} (1,2,\ldots,11)$   \\
$G_2$ & $4$    & $\mathrm{lcm} (1,2,\ldots,5)$    \\
\hline
\end{tabular}
\caption{The dual Coxeter numbers and charge group integer $L$ for the simple %
Lie algebras}
\label{table1}
\end{center}
\end{table}

WZW models also possess D-branes that only preserve the affine symmetry up to
some twist. For every
automorphism of the finite dimensional algebra $\overline{{g}}$, $\omega$-twisted
D-branes can be constructed.
These are solutions of the ``gluing" condition
\begin{equation}
J(z)= \omega \cdot \overline{J} (\overline{z}) \, , z=\overline{z} \, ,
\end{equation}
where $J, \overline{J}$ are the chiral currents of the WZW model.
These D-branes are labelled by the $\omega$-twisted highest weight
representations of $g_k$. The charge group is of the form
$\mathbb{Z}_{M^{\omega}}$, where $M^{\omega}$ is the twisted
analogue of the integer $M$ from the untwisted case. The charge
carried by the D-brane labelled by the $\omega$-twisted highest
weight $a$ has an integer charge $\mathrm{q}_a ^{\omega}$, such
that
\begin{eqnarray}
\mathrm{dim} (\lambda)\, \mathrm{q}_{a}^{\omega} = \sum _ b
\mathcal{N}_{\lambda a} ^{b} \, \mathrm{q}_{b}^{\omega} \quad
\mathrm{mod}\, M^{\omega}\, , \label{twisteqn}
\end{eqnarray}
where $\mathcal{N}_{\lambda a} ^{b}$ are the NIM-rep coefficients that appear in
the Cardy analysis of these D-branes.
$M^{\omega}$ is the largest integer such that (\ref{twisteqn}) holds, again
assuming that all
the charges $\mathrm{q}_a ^{\omega}$ are relatively prime integers. However the
difficulty in carrying over the analysis from the
untwisted case is that there is no brane label $a$ playing the role of the
identity field, and thus we need to resort to a slightly different, and more
complicated,
analysis to determine the charges and $M^{\omega}$.

It was suggested in \cite{FS1,GG1} that the NIM-rep coefficients
$\mathcal{N}_{\lambda a} ^{b}$ are actually the twisted fusion
rules that describe the WZW fusion of the twisted representation
$a$ with the untwisted representation $\lambda$ to give the
twisted representation $b$. Thus the conformal highest weight
spaces of all three representations $\lambda, a$, and $b$ form
representations of the invariant horizontal subalgebra $\overline
g ^{\omega}$ that consists of the $\omega$-invariant elements of
$\overline g$ (For details on such matters, see \cite{FS}). The
twisted fusion rules are a level $k$ truncation of the tensor
product coefficients of the horizontal subalgebra. This
establishes a parallel with the untwisted case, where the
untwisted fusion rules are the level $k$ truncation of the tensor
product coefficients of $\overline g$. Thus by analogy with
(\ref{untchar}), we can make the ansatz
\begin{equation}
\mathrm{q}_a ^{\omega} = \mathrm{dim} _{\overline{g} ^{\omega}}
(a) \, , \label{ansatz}
\end{equation}
i.e. the charge is simply the $\overline{g} ^{\omega}$-Weyl
dimension of the finite part of the twisted weight $a$. Using this
ansatz the integer $M^{\omega}$ was calculated in \cite{GG2} for the
chirality flip twisted $A_n$ and $D_n$ series, and it was also shown
that, up to rescaling, (\ref{ansatz}) is the unique solution to
(\ref{twisteqn}).

The remaining non-trivial cases of triality twisted $D_4$ and charge
conjugation twisted $E_6$ are dealt with in this paper, and require
%a substantial amount of
some nontrivial
Lie theory, especially pertaining to twisted affine Lie algebras. The relevant
background can be found in \cite{GG1,FS,PZ}.

\section{Triality Twisted $D_4$ Brane Charges}
$D_4$ has five non-trivial conjugations, whose NIM-reps can all be
determined from analyzing just the ones corresponding to chirality
flip (which has already be done in \cite{GG1,GG2}) and triality. The latter
is an order three automorphism of the Dynkin diagram $\omega$ that
sends the Dynkin labels $(\lambda_0;\lambda_1,\lambda_2,\lambda_3,\lambda_4)$ to
$(\lambda_0;\lambda_4,\lambda_2,\lambda_1,\lambda_3)$. %
Thus the relevant twisted algebra here is $D_4 ^{(3)}$ with a
horizontal subalgebra $G_2$, labelling $\omega$-invariant states.
Thus boundary states are labelled by triples $(a_0;a_1,a_2)$ where
the level $k=a_0+2a_1+3a_2$. In \cite{GG1}, it is shown how to
express the twisted NIM-reps in terms of $A_2$ fusion rules at
level $k+3$ via the branching $D_4\supset G_2 \supset A_2$:
\begin{eqnarray}
\mathcal {N} _{\lambda a }^{b} = \sum _{i=0} ^{2} \sum _{\gamma ''} b^{\lambda}
_{\gamma ''} \left( N_{J^i \gamma '', a''}^{b''}-N_{J^i \gamma '',
Ca''}^{b''}\right) \, , \label{DAfusion}
\end{eqnarray}
where $C$ denotes charge conjugation in $A_2$, which %
takes a dominant $A_2$ weight to its dual by interchanging the
finite Dynkin labels and $J$ is the simple current of $A_2$ that acts by cyclic
permutation of the Dynkin labels of the $A_2^{(1)}$ weights, and the
$b^{\lambda}_{\gamma ''}$ are the $D_4\supset G_2 \supset A_2$
branching rules (see for example \cite{MP}). The relation between $D_4$ %
boundary states and the weights of $G_2^{(1)}$ and $A_2^{(1)}$ is given by %
the identifications of the appropriate Cartan subalgebras. Explicitly, we %
write \cite{GG1} %
\begin{eqnarray}
a'=\iota (a_0;a_1,a_2)&=&(a_0+a_1+a_2+2;a_2,a_1)\in P_+^{k+2}(G_2)\ ,\label{iota1}\\
a''=\iota'a'=\iota'\iota(a_0;a_1,a_2)&=&(a_0+a_1+a_2+2;a_2,a_1+a_2+1)
\in P_+^{k+3}(A_2)\ .\label{iota2}
\end{eqnarray} %
In the following, level
$k$-$D_4$ quantities (weights and boundary states) are unprimed, while the %
 corresponding level %
$k+2$-$G_2$ weights and level $k+3$-$A_2$ weights are singly and doubly %
primed, respectively. %

Following \cite{GG2}, we make the ansatz that the charge
$\mathrm{q}_a ^{\omega}$ is the $G_2$ Weyl dimension of the
horizontal projection (finite part) of the weight i.e.
\begin{equation}
\mathrm{q}_a ^{\omega} = \mathrm{dim}_{G_2}(a')\, .
\end{equation}
Then for an arbitrary dominant integral weight $\lambda$ of $D_4$ the left hand
side of (\ref{twisteqn}) reads:
\begin{eqnarray}
\mathrm{dim}_{D_4}(\lambda )\, \mathrm{dim}_{G_2} (a') &=& \sum _{\gamma '}
b^{\lambda}_{\gamma '} \, \mathrm{dim}_{G_2} (\gamma ') \, \mathrm{dim}_{G_2} (a')
\nonumber \\
&=&\sum_{\gamma '} b^{\lambda}_{\gamma '} \sum_{b ' \in
P^{k+2}_{+} (G_2)} N_{\gamma ' a'}^{b'} \, \mathrm{dim}_{G_2}
(b ') \qquad \mathrm{mod} \, M_{G_2} \, ,\label{lhs0}
\end{eqnarray}
where $b^{\lambda}_{\gamma '}$ are the $D_4\supset G_2$ branching
rules, and in the second line we have used (\ref{unteqn}) %the charge calculation result
for the untwisted $G_2$ branes at level $k+2$. Now from Table
\ref{table1} we know that at level $k+2$ $M_{G_2}$ is the same as
$M_{D_4}$ at level $k$:
\begin{equation}
M_{G_2}=M_{D_4}=\frac{k+6}{\mathrm{gcd}(k+6,2^2.3.5)} \, ,\label{MD4}
\end{equation}
 and so (\ref{lhs0}) holds %the above statement is true
$\mathrm{mod}\, M_{D_4}$. Now $G_2$ fusion rules at level $k+2$ can
be written in terms of the level $k+3$ fusion rules of $A_2$
following \cite{G1}
\begin{eqnarray}
N_{\gamma ' a'}^{b '}= \sum_{\gamma ''} b^{\gamma '}_{\gamma ''}
\left[N_{\gamma '' a''}^{b ''} - N_{\gamma '' Ca''}^{b ''}\right] \, , \label{Uniqfus}
\end{eqnarray}
where $b^{\gamma '}_{\gamma ''}$ are the $G_2\supset A_2$
branching rules.
%At the level of boundary states, the mapping $D_4\rightarrow G_2\rightarrow A_2$ is given by
%$a=(a_0;a_1,a_2)\rightarrow a'=(a_0+a_1+a_2;a_2,a_1) \rightarrow
%a''=(a_0+a_1+a_2+2;a_2,a_1+a_2+1)$. This is a special ``branching" obtained by identification of the Cartan subalgebras.
%As explained in \cite{GG1} and \cite{G1}, this is obtained using the Verlinde formula by analyzing the
%subset of images of dominant integral weights under the branching. Now
Using this and the fact that $\sum _{\gamma '} b^{\lambda}
_{\gamma'} b^{\gamma '}_{\gamma ''} = b^{\lambda}_{\gamma ''}$, we
rewrite the left hand side of (\ref{twisteqn}) as
\begin{eqnarray}
\mathrm{L.H.S.}=\sum_{\gamma ''} b^{\lambda}_{\gamma ''} \sum_{b ' \in
P^{k+2}_{+} (G_2)}\left[N_{\gamma '' a''}^{b ''} - N_{\gamma '' Ca''}^{b
''}\right] \mathrm{dim}_{G_2} (b ') \qquad \mathrm{mod} \, M_{D_4} \, .
\label{LHS1}
\end{eqnarray}

In order to relate this to the right hand side of (\ref{twisteqn})
where the summation is only over the boundary states of triality
twisted $D_4$, we need to restrict the %above
summation (\ref{LHS1}) somehow to %
the set $\mathcal{D}=Im(\iota'\iota)$ of images $b\mapsto b''$ under
%the mapping $D_4\rightarrow A_2$ described above.
(\ref{iota1}),(\ref{iota2}). %
To do this, we first describe the relevant sets precisely.

An $A_2^{(1)}$ weight $(b_0'';b_1'',b_2'')$ belongs to $\mathcal{D}, J\mathcal{D}$, or
$J^2\mathcal{D}$ respectively, if
\begin{eqnarray}
\mathcal{D} \quad &:& \quad b_0'' > b_2'' >b_1''\geq 0 \, , \nonumber \\
J\mathcal{D} \quad &:& \quad b_1'' > b_0'' >b_2'' \geq 0 \, ,  \\
J^2\mathcal{D} \quad &:& \quad b_2'' > b_1'' >b_0'' \geq 0 \, , \nonumber
\end{eqnarray}
where $J$ is the $A_2$ simple current acting on $A_2^{(1)}$ weights by $J(a_0'';a_1'',a_2'')=(a_2'';a_0'',a_1'')$. %

The set $\mathcal{G}=\iota'(P_+^{k+2}(G_2))$ of images of (\ref{iota2}) %$P^{k+2}_{+} (G_2)$ under the $G_2\supset A_2$ branching
(the set over which we are summing in
(\ref{LHS1})) only has the constraint $b_2''>b_1''\geq 0$. Thus a moment of thought will show that
\begin{equation}
\mathcal{G}=\mathcal{D} \,\cup\, J^2 \mathcal {D}\, \cup\, CJ\mathcal{D}\,\cup
\,\mathcal {B} \, ,
\end{equation}
where %, as before, $C$ is $A_2$ charge conjugation, and
%$\mathcal{B}$ is everything contained in $\mathcal{G}$ not in
%one of the other three sets. A moment of reflection will show that
$\mathcal{B}$ consists of weights in $\mathcal{G}$ such that
either $b_0''=b_1''$ or $b_0''=b_2''$. The following hidden symmetries %facts
are established in the appendices:
\begin{eqnarray}
&&b'\in P^{k+2}(G_2) \Rightarrow\mathrm{dim}_{G_2} (J b')=\mathrm{dim}_{G_2} (J^2 b')=\mathrm{dim}_{G_2} %
(b') \quad \mathrm{mod}\, M_{D_4} \, , \label{D4appfac1}\\
&&b''\in\mathcal{B} \Rightarrow \mathrm{dim}_{G_2} (b')=0 \quad \mathrm{mod}\,
M_{D_4} \, ,\label{D4appfac2} \\
&&b'\in P^{k+2}(G_2) \Rightarrow\mathrm{dim}_{G_2}(Cb')=-\mathrm{dim}_{G_2}(b') \, , \label{D4appfac3} %
\end{eqnarray}
where $C$ and $J$ act on $G_2^{(1)}$ weights through conjugation by $\iota'$:
%formally the action of the $A_2$ conjugation $C$, and the $A_2^{1}$ simple current $J$ on $G_2$ weights are given by
\begin{eqnarray}
C(b_0';b_1',b_2')&=&(b_0';b_1'+b_2'+1,-b_2'-2) \, , \label{G2C}\\
J(b_0';b_1',b_2')&=&(b_1'+b_2'+1;b_0',b_1'-b_0'-1) \,. \label{G2J}
\end{eqnarray}
Here and elsewhere, we write `dim$_{G_2}(a')$' even when $a'$ is not
dominant, by formally evaluating the Weyl dimension formula for $G_2$ at
$a'$. The minus sign in (\ref{D4appfac3})
%has a minus sign reflecting the fact that the image of
indicates that $Cb'$ won't be a dominant $G_2$ weight when $b'$ is --- indeed,
%under $C$ needs to be reflected by an order one member of
$C$ belongs to the Weyl Group of $G_2$.
%to once again be a dominant weight (The minus sign comes from the
%sign/determinant of the Weyl transformation used).

Using these, % and that $J^3=1$, with a little bit of effort
we can rewrite (\ref{LHS1}) as
\begin{eqnarray}
\mathrm{L.H.S.} &=& \sum_{\gamma ''} b^{\lambda}_{\gamma''}\left[ \sum_{b '' %
\in \mathcal {D}}
[N_{\gamma '' a''}^{b ''} - N_{\gamma '' Ca''}^{b''}]\, \mathrm{dim}_{G_2}(b')\right. \nonumber \\
&+& \left. \sum_{b '' \in J^2\mathcal {D}} [N_{\gamma '' a''}^{b ''} -
N_{\gamma '' Ca''}^{b
''}]\, \mathrm{dim}_{G_2}(b')\right. \nonumber \\  %
&+& \left. \sum_{b '' \in J\mathcal {D}} [N_{\gamma ''
a''}^{Cb ''} - N_{\gamma '' Ca''}^{Cb ''}]\,
\mathrm{dim}_{G_2}(Cb')\right] \quad \mathrm{mod}\, M_{D_4} \, ,
\label{LHSD4}
\end{eqnarray}
where we note that there is no contribution from $\mathcal{B}$
due to (\ref{D4appfac2}). Using (\ref{D4appfac3}), the symmetry
$N_{\gamma'' a''}^{\nu''}=N_{C\gamma'' Ca''}^{C\nu''}$, and the
fact that the $D_4\supset A_2$ branching has $b^{\lambda}_{\gamma''}=
b^{\lambda}_{C\gamma''}$ for all $\gamma''$, to simplify the third sum in
(\ref{LHSD4}), we finally obtain
\begin{eqnarray}
\mathrm{L.H.S.} &=& \sum_{i=0}^2\sum_{\gamma ''} b^{\lambda}_{\gamma''}
\sum_{b ''
\in J^i\mathcal {D}}
[N_{\gamma '' a''}^{b ''} - N_{\gamma '' Ca''}^{b''}]\, \mathrm{dim}_{G_2}(b')
 \quad \mathrm{mod}\, M_{D_4} \, .\label{LHSD5}
\end{eqnarray}
But applying (\ref{DAfusion}), (\ref{D4appfac1}), and
the symmetries of fusion rules under the action of simple
currents, we see that (\ref{LHSD5}) also equals the right hand side of
(\ref{twisteqn}).
Thus (\ref{twisteqn}) is indeed satisfied by our ansatz
\begin{eqnarray}
\mathrm{q}_a^{\omega} = \mathrm{dim}_{G_2} (a') \qquad
\mathrm{and}\qquad M^{\omega}=M_{D_4} \, ,
\end{eqnarray}
that is, the charges are given by the Weyl dimension of the representation of
the horizontal subalgebra, and the charge group is the same as in the untwisted
case. As we show in Section 5, the charges are unique up to a rescaling by a
constant factor.

\section{Charge Conjugation Twisted $E_6$ Brane Charges}
$E_6$ has a non-trivial order two symmetry of the Dynkin diagram
that sends the Dynkin labels $(\lambda_0;\lambda_1,\lambda_2,\lambda_3,
\lambda_4,\lambda_5,\lambda_6)$ to %
$(\lambda_0;\lambda_5,\lambda_4,\lambda_3,\lambda_2,\lambda_1,\lambda_6)$. %
The relevant twisted algebra here is $E_6 ^{(2)}$, with a
horizontal subalgebra $F_4$ labelling the $\omega$ invariant
states. The boundary states are labelled by quintuples
$(a_0;a_1,a_2,a_3,a_4)$ such that $k=a_0+2a_1+3a_2+4a_3+2a_4$.
Again, in \cite{GG1} it is shown how to express the twisted
NIM-reps in terms of the untwisted fusion rules of $D_4$ at level
$k+6$ via the branching $E_6\supset F_4\supset D_4$:
\begin{eqnarray}
\mathcal {N} _{\lambda a }^{b} = \sum_{J}\sum _{\pi} \sum _{\gamma ''}
\epsilon (\pi )\, b^{\lambda}
_{\gamma ''} \, N_{J \gamma '', \pi a''}^{b''}\, , \label{EDfusion} %
\end{eqnarray}
where the sum over $\pi$ is over all 6 conjugations of $D_4$
consisting of permutations of the 1st, 3rd, and 4th Dynkin labels,
$b^{\lambda}_{\gamma ''}$ are the $E_6\supset F_4 \supset D_4$
branching rules and $\epsilon(\pi)$ is the parity of the
permutation.
The summation labelled by $J$ is over the four %non-trivial
simple currents of $D_4$: the identity,
$J_{\nu}b''=(b_1'';b_0'',b_2'',b_4'',b_3'')$,
$J_sb''=(b_4'';b_3'',b_2'',b_1'',b_0'')$, and $J_{\nu}
J_s$. Note that each of these simple currents has order 2.
%At the level of
The $E_6$ boundary states
%, the mapping $E_6\rightarrow F_4\rightarrow D_4$ is given by
are related to $F_4^{(1)}$ and $D_4^{(1)}$ weights through the maps \cite{GG1} %
\begin{eqnarray}
a'&=&\iota (a_0;a_1,a_2,a_3,a_4)=(a_0+a_1+a_2+a_3+3;a_4,a_3,a_2,a_1)\in P_+^{k+3}(F_4)\label{iota3}\\
a''&=&\iota'a'=\iota'\iota(a_0;a_1,a_2,a_3,a_4)\label{iota4}\\
&=&(a_0+a_1+a_2+a_3+3;a_1+a_2+a_3+2,a_4,a_3,a_2+a_3+1)\in P_+^{k+6}(D_4)\ .
\nonumber\end{eqnarray} %
%$a=(a_0;a_1,a_2,a_3,a_4)\rightarrow
%a'=(a_0+a_1+a_2+a_3+3;a_4,a_3,a_2,a_1) \rightarrow
%a''=(a_0+a_1+a_2+a_3+3;a_1+a_2+a_3+2,a_4,a_3,a_2+a_3+1)$.
%This is again a special ``branching'' obtained by
These again correspond to the identification of the respective Cartan
subalgebras.
Unprimed quantities refer to level $k$-$E_6$ quantities, while their
corresponding level $k+3$-$F_4$  %quantities are singly primed and the
and level $k+6$-$D_4$ weights are singly and doubly primed, respectively.
%quantities are doubly primed.
Again, as explained in \cite{GG1} and \cite{G1}, these relations are
established by examining the twisted version of the Verlinde formula.

Following \cite{GG2} again, we take the ansatz that the charge
$\mathrm{q}_a ^{\omega}$ is the $F_4$ Weyl dimension of the
finite part of the weight i.e.
\begin{equation}
\mathrm{q}_a ^{\omega} = \mathrm{dim}_{F_4}(a')\, .
\end{equation}
Then for an arbitrary dominant integral weight $\lambda$ of $E_6$
the left hand side of (\ref{twisteqn}) reads:
\begin{eqnarray}
\mathrm{dim}_{E_6}(\lambda ) \, \mathrm{dim}_{F_4} (a') &=& \sum
_{\gamma '} b^{\lambda}_{\gamma '}\, \mathrm{dim}_{F_4} (\gamma ')
\,\mathrm{dim}_{F_4} (a')
\nonumber \\
&=&\sum_{\gamma '} b^{\lambda}_{\gamma '} \sum_{b ' \in
P^{k+3}_{+} (F_4)} N_{\gamma ' a'}^{b'} \, \mathrm{dim}_{F_4}
(b ') \qquad \mathrm{mod} \, M_{F_4} \, ,\label{lhs00}
\end{eqnarray}
where $b^{\lambda}_{\gamma '}$ are the $E_6\supset F_4$ %$F_4\supset D_4$
branching rules, and in the second line we have used %the charge calculation result
(\ref{unteqn}) for the untwisted $F_4$ branes at level $k+3$. From Table
\ref{table1} we know that at level $k+3$ $M_{F_4}$ is the same as
$M_{E_6}$ at level $k$:
\begin{equation}
M_{F_4}=M_{E_6}=\frac{k+12}{\mathrm{gcd}(k+12,2^3.3^2.5.7.11)} \, ,\label{ME6} %
\end{equation}
 and so (\ref{lhs00}) %the above statement is true
also holds $\mathrm{mod}\, M_{E_6}$.

Now $F_4$ fusion rules at level $k+3$ can be written in terms of
the level $k+6$ fusion rules of $D_4$ following \cite{G1}
\begin{eqnarray}
N_{\gamma ' a'}^{b '}= \sum_{\gamma ''} \sum_{\pi} \epsilon (\pi
) \,b^{\gamma '}_{\gamma ''} \,N_{\gamma '' \pi a''}^{b ''}  \, ,%
\end{eqnarray}
where $b^{\gamma '}_{\gamma ''}$ are the $F_4\supset D_4$
branching rules, and the $\pi\in S_3$ are as before.
%the $D_4$ conjugations obtained by permuting the 1st, 3rd, and 4th Dynkin labels.
As explained in \cite{GG1} and \cite{G1}, this is obtained using the Verlinde
formula by analyzing the subset of images of dominant integral
weights under the branching. Now using this and the fact that
$\sum _{\gamma '} b^{\lambda} _{\gamma'} b^{\gamma '}_{\gamma ''}
= b^{\lambda}_{\gamma ''}$, we rewrite the left hand side of
(\ref{twisteqn}) as
\begin{eqnarray}
\mathrm{L.H.S.}=\sum_{\gamma ''} \sum_{b ' \in P^{k+3}_{+}
(F_4)} \sum_{\pi} \epsilon(\pi) \, b^{\lambda}_{\gamma ''}\,
N_{\gamma '' \pi a''}^{b ''} \, \mathrm{dim}_{F_4} (b ')
\qquad \mathrm{mod} \, M_{E_6} \, . \label{E6LHS1} %
\end{eqnarray}

In order to relate this to the right hand side of (\ref{twisteqn})
where the summation is only over the boundary states of twisted
$E_6$, we need to restrict the above summation somehow to the set
$\mathcal{E}=Im(\iota'\iota)$ of images $b\mapsto b''$.
% under the mapping $F_4\rightarrow D_4$.
To do this, we first describe the
relevant sets precisely.

A $D_4^{(1)}$ weight $(b_0'';b_1'',b_2'',b_3'',b_4'')$ belongs to $\mathcal{E},
J_{\nu}\mathcal{E}$, $J_s\mathcal{E}$, or $J_{\nu}
J_s\mathcal{E}$, where the $J$ are the $D_4$ simple currents, if
\begin{eqnarray}
\mathcal{E} &:& b_0'' > b_1'' > b_4''>b_3''\geq 0\, , \nonumber \\
J_{\nu}\mathcal{E} &:& b_1'' > b_0'' > b_3''>b_4''\geq 0 \, ,\nonumber \\
J_s\mathcal{E} &:& b_4'' > b_3'' > b_0''>b_1''\geq 0 \, , \\
J_{\nu}  J_s\mathcal{E} &:& b_3'' > b_4'' > b_1''>b_0''\geq 0 \, .\nonumber
\end{eqnarray}
The set $\mathcal{F}$ of images of $P^{k+3}_{+} (F_4)$ under the
$F_4\supset D_4$ branching (the set over which we are summing in
(\ref{E6LHS1}) ) only has the constraints $b_1'' > b_4''>b_3''\geq 0$. Thus a moment of thought will show that
\begin{equation}
\mathcal{F}=\mathcal{E}\, \cup\, \pi _{143} J_{\nu} \mathcal {E}\, \cup \,\pi
_{341} J_s\mathcal{E}\, \cup \,\pi _{413} J_{\nu}  J_s\mathcal{E}\,
\cup\, \mathcal {B} \, ,
\end{equation}
where $\pi_{abc}$ is the $D_4$ conjugation taking the Dynkin labels 1,3 and 4
respectively to $a,b$ and $c$, and $\mathcal{B}$
%is everything contained in $\mathcal{F}$ not in one of the other four sets. A
%moment of reflection will show that $\mathcal{B}$
consists of weights in $\mathcal{F}$ such that either
$b_0''=b_1''$ or $b_0''=b_2''$ or $b_0''=b_4''$.

The following facts, where the $\pi$ are the $D_4$ conjugations and the $J$
are any of the $D_4$ simple currents, are proved in the appendices:
\begin{eqnarray}
&&\mathrm{dim}_{F_4} (\pi b') = \epsilon (\pi) \, \mathrm{dim}_{F_4} (b') \quad \forall b'\in P^{k+3}(F_4)\, , \label{E6appfac1} \\ %
&&\mathrm{dim}_{F_4} (Jb')= \mathrm{dim}_{F_4} (b') \quad \mathrm{mod} \, M_{F_4} \quad \forall b' \in P^{k+3}(F_4)\label{E6appfac2} \, , \\
&&\mathrm{dim}_{F_4} (b')=0 \quad \mathrm{mod}\, M_{E_6} \quad \forall b''\in\mathcal{B}\, . \label{E6appfac3}
\end{eqnarray}
The action of the $D_4$ conjugations $\pi$ and simple currents $J$ on $F_4^{(1)}$
weights can be easily obtained by converting $F_4^{(1)}$ weights to $D_4^{(1)}$ weights
using $\iota'$, applying $\pi$ or $J$, and then converting back to $F_4^{(1)}$
using $\iota'{}^{-1}(d_0'';d_1'',d_2'',d_3'',d_4'')=
(d_0'';d_2'',d_3'',d_4''-d_3''-1,d_1''-d_4''-1)$.
As for $G_2$, we write `dim$_{F_4}(a')$' even when $a'$ is not
dominant, by formally applying the Weyl dimension formula.
The factor $\epsilon (\pi)$ in (\ref{E6appfac1}) is the parity $\pm 1$, and
as before, each $\pi$ belongs to the Weyl Group of $F_4$.
%considering the inversion of the map $F_4\rightarrow D_4: (f_0';f_1',f_2',f_3',f_4')\rightarrow (f_0';f_2'+f_3'+f_4'+2,f_1',f_2',f_2'+f_3'+1)$ which is given by
%$D_4\rightarrow F-4: (d_0'';d_1'',d_2'',d_3'',d_4'')\rightarrow(d_0'';d_2'',d_3'',d_4''-d_3''-1,d_1''-d_4''-1)$.
%As in the $G_2$ case, the $\epsilon (\pi)$ factor in the dimensional relation (\ref{E6appfac1}) refers to the order of the
%element of the Weyl group needed to reflect the image weight into the fundamental Weyl chamber.

Using these, % and the fact that all of the $D_4$ simple currents square to the
%identity, with a little bit of effort
we can rewrite (\ref{E6LHS1}) as
\begin{eqnarray}
\mathrm{L.H.S.} &=& \sum_{\gamma ''} b^{\lambda}_{\gamma ''}\sum_{\pi} %
\epsilon(\pi)\,\left[\sum_{b '' \in \mathcal {E}} %
N_{\gamma '' \pi a''}^{b ''} \, \mathrm{dim}_{F_4}(b') \right.
\nonumber \\
&+& \left. \sum_{b '' \in
J_{\nu}\mathcal {E}}
N_{\gamma '' \pi a''}^{\pi_{143}b ''} \, \mathrm{dim}_{F_4}(\pi_{143}b') %
+ \sum_{b '' \in J_{s}\mathcal {E}} %
N_{\gamma '' \pi a''}^{\pi_{341}b ''} \, \mathrm{dim}_{F_4}(\pi_{341}b') %
\right.\nonumber \\
&+& \left. \sum_{b '' \in J_{\nu}  J_s \mathcal {E}}
N_{\gamma '' \pi a''}^{\pi_{413} b ''} \, %
\mathrm{dim}_{F_4}(\pi_{413} b') \right] \quad \mathrm{mod}\, M_{E_6} %
\, ,\label{LHSE6}
\end{eqnarray}
where we note that there is no contribution from $\mathcal{B}$
due to (\ref{E6appfac3}). Exactly as for the argument last section, the
symmetries of the fusion rules under $\pi$ and simple currents, the symmetry
$b^\lambda_{\pi \gamma''}=b^\lambda_{\gamma''}$ of the branching rules, together
with the hidden symmetries (\ref{E6appfac1}),(\ref{E6appfac2}) and the
expression (\ref{EDfusion}), show that
\begin{eqnarray}
\mathrm{L.H.S.} &=& \sum_{\gamma ''} b^{\lambda}_{\gamma ''}\sum_J\sum_{\pi} %
\epsilon(\pi)\,\sum_{b '' \in \mathcal {E}} %
N_{J\gamma '' \pi a''}^{b ''} \, \mathrm{dim}_{F_4}(b')
=\mathrm{R.H.S.} \quad \mathrm{mod}\, M_{E_6} %
\, .
\end{eqnarray}
%Now, using (\ref{E6appfac1}) and (\ref{E6appfac2}) and the properties of fusion rules under simple
%currents, we rewrite as
%\begin{eqnarray}
%\mathrm {L.H.S} =\sum_{\gamma ''} b^{\lambda}_{\gamma ''}
%\sum_{J}\sum_{\pi}\sum_{b''\in \mathcal{E}} \epsilon({\pi}) \, N _{J\gamma
%'' \pi a''} ^{b''} \, \mathrm{dim}_{F_4}(b') \quad \mathrm{mod}\, M_{E_6}
%\,,
%\end{eqnarray}
%which upon referring to (\ref{EDfusion}), and noting that $\mathcal{E}$ is the
%set of all images of the boundary states under the $E_6\rightarrow D_4$
%mapping, can be written as
%\begin{eqnarray}
%\mathrm {L.H.S}=\sum_{b} \mathcal{N}_{\lambda a}^{b} \, \mathrm{dim}_{F_4}(b')
%\quad \mathrm{mod}\, M_{E_6} \, ,
%\end{eqnarray}
%and this is precisely the right hand side of (\ref{twisteqn}). Q.E.D.
Thus, again, (\ref{twisteqn}) is indeed satisfied by our ansatz
\begin{eqnarray}
\mathrm{q}_a^{\omega} = \mathrm{dim}_{F_4} (a') \qquad
\mathrm{and}\qquad M^{\omega}=M_{E_6} \, ,
\end{eqnarray}
that is, the charges are once again given by the Weyl dimension of the
representation of
the horizontal subalgebra, and the charge group is the same as in the untwisted
case. As we show next, the charges are unique up to a rescaling by a constant
factor.

\section{Uniqueness}
We need to show that the solutions found to the charge equation (\ref{twisteqn}) in both the $D_4$ and $E_6$ cases are unique up
to rescaling. To this end it is sufficient to prove that if the charge equation is satisfied by a set of integers
$\tilde{q}_a$ modulo some integer $\tilde{M}$, then
\begin{eqnarray}
\tilde{q}_a=\mathrm{dim}(a') \, \tilde {q}_0 \quad \mathrm{mod} \, \tilde{M} \,.\label{prop}
\end{eqnarray}
In this case, we can divide all charges by $\tilde{q}_0$, and the charge
equation will still be satisfied if we also divide $\tilde{M}$
by $\mathrm{gcd}(\tilde{q}_0,\tilde{M})$.
Finally, by an argument due to Fredenhagen  \cite{GG2}\ we get that $M':=
\tilde{M}/\mathrm{gcd}(\tilde{q}_0,\tilde{M})$ must divide our $M$.
Explicitly, by construction, $M$ is the g.c.d.\ of
the dimensions of the elements of the fusion ideal that quotients the
representation ring in
order to obtain the fusion ring. Since NIM-reps provide representations of the
fusion ring, any element of the
fusion ideal acts trivially i.e. $\mathrm{dim(\alpha)}\,\mathrm{dim}(a) =0\,
\mathrm{mod} \, M'$ for any
$\alpha$ in the fusion ideal. Thus, using the fact that the $\mathrm{dim}(a)$
 are relatively prime integers, we
see that $M'$ must divide $M$.  Thus any alternate solution $\tilde{q}_a,
\tilde{M}$ to  (\ref{twisteqn}) which obeys (\ref{prop}), is just a rescaled
version of our ``standard'' one $\mathrm{q}_a,M$.

We will work with $D_4$, the proof for $E_6$ is similar and will be sketched at the end.
The $G_2\subset D_4$ branching rules can be inverted: we can formally write
\begin{eqnarray}
a'=\sum_\lambda \tilde{b}_\lambda^a\lambda \, ,\label{invbra}
\end{eqnarray}
where $\tilde{b}_\lambda^a$ are integers (possibly negative),
$a'\in P_+(G_2)$, and the sum is over $D_4$ weights $\lambda$
\cite{G1}. More precisely, (\ref{invbra}) holds at the level of
characters, where the domain of the $D_4$ ones is restricted to
the $\omega$-invariant vectors in the $D_4$ Cartan subalgebra, and
the $G_2$ characters are evaluated at the image of those vectors
by $\iota$. To prove (\ref{invbra}), it suffices to verify it for
the $G_2$ fundamental weights, where we find
$(1,0)=(0,1,0,0)-(1,0,0,0)+(0,0,0,0)$ and
$(0,1)=(1,0,0,0)-(0,0,0,0)$. Since all other $G_2$ weights can be
constructed from the fundamental ones by tensor products, every
dominant $G_2$ weight can formally be inverted under the branching
and written in terms of linear combinations of dominant integral
$D_4$ weights. Then \footnote{In the following, we use the fact
that both (\ref{twisteqn}) and (\ref{DAfusion}) (and
(\ref{EDfusion}) in the case of $E_6$) remain true for all
dominant weights $\lambda$ and not just affine ones. These
expressions were obtained using ratios of S-matrices (see
\cite{GG1}), which can be interpreted as Lie algebra characters in
the case of finite weights, and thus the NIM-reps $\mathcal{N}$
can be continued to include all dominant weights. This
continuation also removes any subtleties in the comparison to
K-theory by evaluating the charge constraint equation for all
dominant weights $\lambda$. We thus can also see there are no
exceptional situations at low levels.}
\begin{eqnarray}
\mathrm {dim}_{G_2} (a')\, \tilde{q}_0&=&\sum_\lambda \tilde{b}_\lambda^a \,
\mathrm{dim}_{D_4} (\lambda)\, \tilde {q}_0 \nonumber \\
&=& \sum_{\lambda} \tilde{b}_\lambda^a  \sum_{b'} \mathcal{N}^{b}_{\lambda,0} \tilde{q}_b \quad \mathrm{mod} \, \tilde{M} \, ,\label{uniq}
\end{eqnarray}
where we have used the charge equation, which the $\tilde{q}_a$ satisfy modulo $\tilde{M}$ by assumption. Now
we use the expression (\ref{DAfusion}) to write (\ref{uniq}) in terms of $A_2$ fusions: we get
\begin{eqnarray}
\mathrm{R.H.S.}=\sum_\lambda \tilde{b}_\lambda^a  \sum_{b} \sum_{\gamma''} \sum_{j=0}^2 b^{\lambda}_{\gamma ''} \left[ N^{b''}_{J^j\gamma'',0''}-N^{b''}_{J^j\gamma'',C0''}\right] \tilde{q}_b \, .
\end{eqnarray}
Now however, from (\ref{Uniqfus}) and \cite{G1}, we can express this in terms of $G_2$ untwisted fusion rules, and using properties of the $A_2$ fusion rules under simple currents along the way, we obtain
\begin{eqnarray}
\mathrm{R.H.S.}=\sum_\lambda \tilde{b}_\lambda^a  \sum_{b} \sum_{\gamma'} \sum_{j=0}^2 b^{\lambda}_{\gamma '} N^{J^j b'}_{\gamma' 0'} \, \tilde{q}_{b} \quad \mathrm{mod}\, \tilde{M}\ .
\end{eqnarray}
Note that $0'$ here is the $G_2$ vacuum (whereas $0''$, the image of $0'$ under $\iota'$, is not the $A_2$ vacuum), and thus
\begin{eqnarray}
\mathrm{R.H.S.}=\sum_b \sum_{j=0}^2 \delta _{a'} ^{J^j b'} \, \tilde{q}_{b}=\tilde{q}_a \quad \mathrm{mod} \, \tilde{M} \ ,
\end{eqnarray}
where we have used the facts that $b'$ is never fixed by $J$ and that the sets $\mathcal{D}$ and $J\mathcal{D}$ are disjoint(see (\ref{G2J})).
 Thus (\ref{prop}), and with it uniqueness, is established.

The proof for $E_6$ is virtually identical, except now we use the expression for untwisted $F_4$ fusion rules in terms of $D_4$ fusion rules found in \cite{G1}.

\section{Conclusion}
In this paper, we have shown that the charge groups of the
triality twisted $D_4$ and the charge conjugation twisted $E_6$
branes are identical to those of the untwisted D-branes. This is
in nice agreement with the K-theoretic calculation \cite{VB} and
completes the exceptional cases not dealt with in \cite{GG2}. Our
calculations show that the charges of these twisted D-branes
corresponding to the twisted representation $a$ is the dimension
of the highest weight space of the representation $a$. Thus from
the string theoretic point of view, analogous to the situation
with untwisted D-branes, the charge associated to the D-brane is
the multiplicity of the ground state of the open string stretched
between the fundamental $D0$-brane and the brane labelled by $a$
in question. So, in the supersymmetric version of WZW models, the
charge may be interpreted as an intersection index, motivating
possible geometric interpretation of these results. The explicit
computation of the charges is missing from the K-theoretic
calculations, and has been supplied here.\footnote{During the
preparation of this manuscript the work of \cite{FG} has come to
our attention, in which similar results are derived using
different methodology.} There are no additional unproven
conjectures made in this paper. All the arguments have been proved
up to some conjectures needed from the untwisted cases (i.e. the
content of Table \ref{table1} for $D_4$, $G_2$, $E_6$, and $F_4$).

A number of non-trivial, and somewhat surprising Lie theoretic
identities have been proved along the way. Some of the dimension
formulae regarding the action of simple currents of a subalgebra on
weights of the larger algebra indicate that there might exist
interesting constraints on the larger algebra due to the underlying
symmetry in the branchings. In some sense, the enlarged algebra
``breaks" symmetries of the smaller algebra, but still ``sees" the
underlying symmetry (analogous to ideas of renormalization of
quantum field theories with spontaneously broken symmetries.)

The Lie theoretic meaning of (\ref{D4appfac3}) and (\ref{E6appfac1}) is clear: the $A_2$ and $D_4$ conjugations $C$ and $\pi\in S_3$ are
elements of the Weyl groups of $G_2$ and $F_4$ respectively. The meaning of
(\ref{D4appfac1}) and (\ref{E6appfac2}) is far less clear (though it has to
do with the theory of equal rank subalgebras \cite{GVW}), but it does suggest a far-reaching generalization whenever the Lie algebras share the same Cartan subalgebras --- for example, $A_1\oplus \cdots\oplus  A_1$ ($n$ copies) and $C_n$, or $A_8$ and $E_8$.
Given any simple current $J$ of any affine Lie algebra $g$ at level $k$, it is already surprising that Weyl dimensions for the horizontal subalgebra $\overline{g}$ see the action of
$J$ via $\mathrm{dim}_{\overline{g}}(J\lambda)=\pm \mathrm{dim}_{\overline{g}} (\lambda) \quad \mathrm{mod}\, M_{g_k}$. Far more surprising is that, at least sometimes, if two Lie algebras $\overline{g}$ and $\overline{g}'$ share the same
Cartan subalgebras, then the Weyl dimensions of the first sees the simple currents $J'$ of the second: $\mathrm{dim}_{\overline{g}}(J'\lambda)=\pm \mathrm{dim}_{\overline{g}} (\lambda) \quad \mathrm{mod}\, M_{g_k}$.

\section*{Acknowledgements} We warmly thank Stefan Fredenhagen, Matthias
Gaberdiel, and Mark Walton for valuable exchanges. This research is
supported in part by NSERC.

\section{Appendices}

\newtheorem {lemma} {Lemma}
\newtheorem {theorem} {Theorem}
\newtheorem{fact} {Fact}
We will use the following fact for proofs in both the $D_4$ and $E_6$ cases.

\begin{fact} Suppose $K,L,N$ are arbitrary integers. Write
$M=\frac{K}{\mathrm{gcd}(K,L)}$,
and let $L=\prod_p p^{\lambda_p}$ and $N=\prod_p p^{\nu_p}$ be
prime decompositions. Suppose we have integers $f_i,d_i$ such that both
$\prod_i f_i$ and $\prod_i(f_i-d_iK)$ are divisible by N.
Then
\begin{eqnarray}
\frac{\prod_i f_i}{N}=\frac{\prod_i (f_i-d_i K)}{N} \quad \mathrm{mod} \, M\ ,
\end{eqnarray}
provided that for each prime $p$ dividing $M$ such that $\lambda_p<\nu_p$,
it is possible to find $0\leq\alpha_{i,p}\leq \lambda_p$, such that
$p^{\alpha_{i,p}}$ divides $f_i$ for each $i$, and $\sum_i \alpha_{i,p}
\ge\nu_p$.
\end{fact}

%\begin{fact}
%(a) Suppose $K,L,N$ are arbitrary integers. Write $M=\frac{K}{\mathrm{gcd}(K,L)}$,
%and let $L=\prod_p p^{\lambda _p}$ and $N=\prod_p p^{\nu _p}$ be
%prime decompositions. Then
%\begin{eqnarray}
%\frac{\prod_i a_i}{N}=\frac{\prod_i (a_i-\delta _i K)}{N} \quad \mathrm{mod} \, M\ ,
%\end{eqnarray}
%for arbitrary integers $a_i,\delta _i$, provided that for each prime $p$ dividing $M$,
%it is possible to find $0\leq \alpha _{i,p}\leq \lambda _p$, such that $p^{\alpha _{i,p}}$ divides $a_i$ and $\sum \alpha _{i,p}\ge\nu_p$.
%
%(b) Keep the notation as in part (a). If $\nu_p\le \lambda_p$ for some fixed prime
%$p$, and $\prod_ia_i$ is divisible by $p^{\nu_p}$, then for that $p$ it will
%be automatic that such numbers $\alpha_{i,p}$ can be found.
%\end{fact}

The reason we can restrict to primes $p$ dividing $M$ is that $p$ coprime to
$M$ are invertible modulo $M$, and so can be freely cancelled
on both sides and ignored. For primes $p$ dividing $M$, $a_i/p^{\alpha_{i,p}}
=(a_i-\delta_i K)/p^{\alpha_{i,p}}$ holds $\mathrm{mod}\, M$, $\forall \, i$.
If $\lambda_p\ge\nu_p$, choose $\alpha_{i,p}$ to be the exact power of $p$
dividing $a_i$.
The divisibility by $N$ hypothesis will be automatically satisfied, because
the products we will be interested in come from the Weyl dimension formula.

\subsection{\textbf{Appendix A: $D_4$ Dimension Formulae}}

For any integral weight $a''=(a_0;a_1,a_2)\in P^{k+3}(A_2)$, we can
substitute $\iota'{}^{-1}(a_0;a_1,a_2)=(a_0;a_1,a_2-a_1-1)=a'$ into
the Weyl dimension
formula \cite{FH}\ of $G_2$, in order to express $G_2$-dimensions using $A_2$ Dynkin
labels: %for all weights $a''$ in
%$\mathcal{G}$, the Weyl dimension formula \cite{FH}, we can write
\begin{eqnarray}
\mathrm{dim}_{G_2}(a')=\frac{1}{120}(a_2-a_1)(a_2+1)(2a_2+a_1+3)(a_2+a_1+2)(a_2+2a_1+3)(a_1+1) \label{G2A2dim} \, .%
\end{eqnarray}

\begin{theorem}
$\forall\, a'\in P^{k+2}(G_2),\
\mathrm{dim}_{G_2}(Ca')=-\mathrm{dim}_{G_2}(a')$.
\end{theorem}
This is an automatic consequence of the $a_1\leftrightarrow a_2$ %
anti-symmetry of (7.2). In fact, $C$ is in the Weyl group of $G_2$ %
and so more generally Theorem 1 follows from the anti-symmetry of the
Weyl dimension formula under Weyl group elements.

\begin{theorem} {$\mathrm{dim}_{G_2}(a')=\mathrm{dim}_{G_2}(Ja')=\mathrm{dim}_{G_2}(J^2a') \quad
\mathrm{mod}\ M_{G_2}\quad \forall a' \in P^{k+2}(G_2)$.}
\end{theorem}
\textbf{Proof}:
Using %the Weyl dimension formula for $G_2$ weights
(\ref{G2A2dim}) and $\iota'$, %examining the mapping of boundary states $G_2\rightarrow A_2$ we can show that
we get
\begin{eqnarray}
\mathrm{dim}_{G_2}(J^2a')%&=&\frac{1}{120} (a_1+a_2+2)(a_0+a_2+a_2+3)(a_0+2a_1+2a_2+4) \nonumber \\
%&&*(a_0+1)(2a_0+3a_1+3a_2+8)(a_0+3a_1+3a_2+7) \nonumber \\
&=&\frac{1}{120} (a_1+a_2+2)(a_1+2a_2+3-K)(a_2+1-K)\nonumber \\ %
&&*(2a_1+3a_2+5-K)(a_1+3a_2+4-2K)(a_1+1+K)\nonumber \, ,%
\end{eqnarray}
where we put $K=k+6$ and used the fact that $k=a_0+2a_1+3a_2$. %
In the notation of Fact 1, here $N=120=2^3.3.5$, $L=60=2^2.3.5$. From
Fact 1, it suffices to consider the primes $p$ with $\nu_p>\lambda_p$, i.e.\
$p=2$. To show that
$p=2$ always satisfies the condition of Fact 1, i.e. that the $\alpha _{i,2}$ can be found for any choice of $a_i$, it suffices to verify it
separately for the 16 possible values of $a_1,a_2 \quad \mathrm{mod} \, 2^2$.
Though perhaps too tedious to check by hand, a computer does it in no time.
The proof for dim$_{G_2}(Ja')$ is now automatic from Theorem 1.

\begin{theorem}
Given any $b'\in P^{k+2}(G_2)$, if $Cb'=J^ib'$ for some $i$, then
%$b''\in\mathcal{B} \Rightarrow
$\mathrm{dim}_{G_2} (b')=0$ $\mathrm{mod}\,M_{G_2}$.
\end{theorem}
\textbf{Proof:}
%Consider $a''$ at level $k+3$ i.e. $a'$ at level $k+2$. First
Write $b''=(b_0;b_1,b_2)\in P^{k+3}(A_2)$. By Theorem 1, it suffices to consider the case
where $b_0=b_2$. In this case we can write $k+3=b_1+2b_2$. Thus, again using
%the Weyl dimension formula
(\ref{G2A2dim}) we have
\begin{eqnarray}
\mathrm{dim}_{G_2}(b')&=&\frac{1}{120} (b_1+1)(b_2+1)(b_1+b_2+2)% \nonumber \\
%&&
(3b_2+3-K)K(3b_2+3-2K)\ . \nonumber
\end{eqnarray}
The proof now proceeds as in Theorem 2.

Of course given any weight $b''\in \mathcal{B}$,
$b'=\iota'{}^{-1}(b'')$ will obey the hypothesis of Theorem 3, and so
(\ref{D4appfac2}) follows. Note that combining Theorems 1 and 2,
we get that any weight $b'$ as in Theorem 3 will obey dim$_{G_2}(b')=-\mathrm{dim}_{G_2}(b')$
mod $M_{G_2}$. Thus Theorems 1 and 2 are almost enough to directly get
Theorem 3 (and in fact imply it for all primes $p\ne 2$).

\subsection{\textbf{Appendix B: $E_6$ Dimension Formulae}}

As before, use $\iota'$ and  the Weyl dimension formula for $F_4$ to
write the (formal) Weyl dimension of an arbitrary $F_4$ integral weight
$b'\in P(F_4)$ in terms of the Dynkin labels of the $D_4$ weight
$b''=\iota'(b')=(b_0'';b_1'',b_2'',b_3'',b_4'')$. For convenience write
$a_i=b_i''+1$. Then we obtain
%, using the Weyl dimension formula, the $F_4$ representation of
%highest weight $f$ has dimension
\begin{eqnarray}
&&\frac{1}{2^{15}3^7 5^4 7^2 11}a_1a_2a_3a_4(a_1+a_2)(a_1+a_3)(a_1+a_4)
(a_2+a_3)(a_2+a_4)(a_3+a_4) \nonumber \\ &*&
(a_1-a_3)(a_1-a_4)(a_4-a_3) (a_1+a_2+a_3)(a_1+a_2+a_4)(a_2+a_3+a_4) \label{F4dim}
\\ &*& (a_1+a_2+a_3+a_4)(a_1+2a_2+a_3)(a_1+2a_2+a_4)(2a_2+a_3+a_4)
\nonumber \\ &*& (a_1+2a_2+a_3+a_4) (2a_1+2a_2+a_3+a_4)(a_1+2a_2+2a_3+a_4)
(a_1+2a_2+a_3+2a_4) \,.\nonumber
\end{eqnarray}

\begin{theorem} For any $F_4$ weight $b'\in P(F_4)$ and any outer automorphism
$\pi\in S_3$ of $D_4$,
\begin{eqnarray}
\mathrm{dim}_{F_4} (\pi b') = \epsilon (\pi) \, \mathrm{dim}_{F_4} (b') \quad
%\forall b''\in\mathcal{F}
\, .
\end{eqnarray}
%where the $\pi$ are the conjugations of $D_4$.
\end{theorem}
This follows easily from the Weyl dimension formula (7.3) by explicitly using the action of the $\pi$ on the weights. As with Theorem 1, it expresses
the anti-symmetry of Weyl dimensions under the Weyl group.

\begin{theorem} For all $F_4^{(1)}$ weights $b'\in P^{k+3}(F_4)$,
\begin{eqnarray}
\mathrm{dim}_{F_4} (b')= \mathrm{dim}_{F_4} (J_{\nu}b')=\mathrm{dim}_{F_4} (J_{s}b') %\nonumber \\
=\mathrm{dim}_{F_4} (J_{\nu}  J_s b') \quad \mathrm{mod}\, M_{F_4}\, .
% \quad \forall b''\in\mathcal{F} \, .
\end{eqnarray}
\end{theorem}
%\textbf{Proof:}
%We will indicate the proof for $J_{\nu}$, the remaining simple currents can be obtained by using the various $D_4$ conjugations.
From Theorem 4, it suffices to consider only $J_\nu$.
%From Table \ref{table1}, we see that for $F_4$ at level $k+3$, $M$ is of the form $\frac{K}{\mathrm{gcd}(K,L)}$, where $K=k+12$ and $L=\mathrm{lcm}(1,...,11)=2^3.3^2.5.7.11$.
%Given a $D_4$ weight $b''=(b_0,b_1,b_2,b_3,b_4)$, the $F_4\supset D_4$ branching can be inverted to give the $F_4$ weight $b'=(b_0,b_2,b_3,b_4-b_3-1,b_1-b_4-1)$ (assuming of course that $b''\in \mathcal{F}$). We relabel the terms in the $F_4$ dimension formula for the weight $b'$ as
%\begin{eqnarray}
%\mathrm{dim}_{F_4}(b')=\frac{\prod _{i=1}^{24} d_i}{W} \, ,
%\end{eqnarray}
%such that
%\begin{eqnarray}
%\mathrm{dim}_{F_4}(\pi_{143} J_{\nu}b')=\frac{\prod _{i=1}^{24} (d_i-\delta _i K)}{W} \, ,
%\end{eqnarray}
%where $\delta _i$ is $0$ if $1\leq i\leq 9$, $1$ if $10\leq i \leq 23$ and $2$ if $i=24$.
%
The proof uses Fact 1 and an easy computer check for primes $p=2,3,5,7$, %
as in the proof of Theorem 2. %

\begin{theorem} Let $b'\in P^{k+3}(F_4)$ be any $F_4^{(1)}$ weight satisfying
$\pi b'=Jb'$, for any $D_4$ simple current $J$ and any order-2 outer automorphism
$\pi$ of $D_4$. Then
%\begin{eqnarray}
%b''\in\mathcal{B} \Rightarrow
$\mathrm{dim}_{F_4} (b')=0 \quad \mathrm{mod}\, M_{F_4} \, .$
%\end{eqnarray}
\end{theorem}
\textbf{Proof}:
The proof of this follows automatically from Theorem 4.
Analogous to the situation for $G_2$, one of the factors in the dimension formula turns out to be $K$. So, accommodating the denominators as in Fact 1
will yield the term
$\frac{K}{p^\alpha}$ for some $0\leq \alpha \leq \lambda_p$. But this is $0$
modulo $M_{F_4}$, for every prime $p$ dividing $M_{F_4}$. Q.E.D.

\subsection{\textbf{Appendix C: NIM-Reps and Graphs}}

In this appendix we give explicit descriptions of some of the
NIM-reps at low level for both $D_4$ and $E_6$ \cite{GG1}. The
NIM-rep graphs characterizing the matrix associated to the %
field $\lambda=\Lambda_1$ are given. The corresponding graph has vertices
labeled by the rows (or columns) of $\mathcal{N}_\lambda$, and the
vertex associated to $i$ and $j$ are linked by
$(\mathcal{N}_{\lambda})_{ij}$ lines.

\begin{figure}
\begin{center}
\includegraphics[width=0.75\textwidth, angle=0]{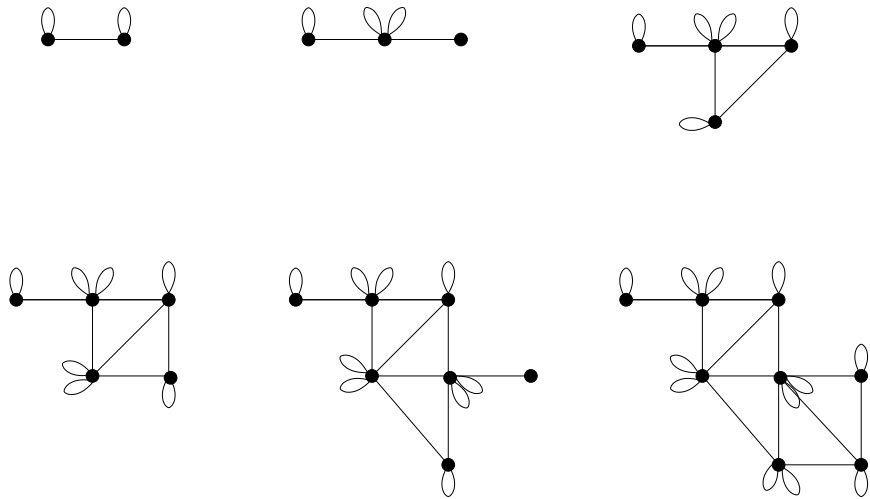}
\caption{NIM-Reps for $D_4$ with Triality $k=2,...,7$} \label{fig1}
\end{center}
\end{figure}

\begin{figure}
\begin{center}
\includegraphics[width=0.75\textwidth, angle=0]{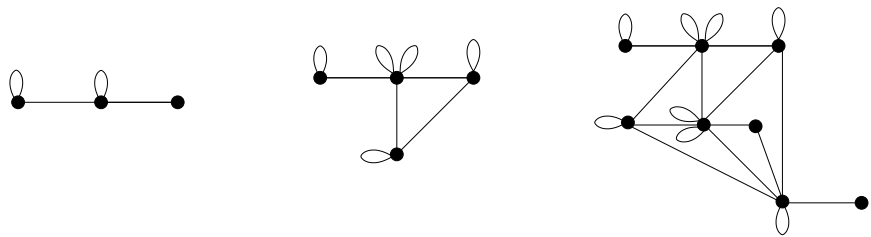}
\caption{NIM-Reps for $E_6$ with Charge Conjugation $k=2,3,4$}
\label{fig2}
\end{center}
\end{figure}

\subsection{\textbf{Appendix D: A Sample Calculation}}

A suitable example to illustrate the situations considered is at
level $k=5$ for triality twisted $D_4$. In this case from Table
\ref{table1}, we get $M_{D_4}=11$. The boundary states are
labelled by triples $(a_0;a_1,a_2)$ such that $k=a_0+2a_1+3a_2$.
The boundary weights then are
\begin{eqnarray}
[5,0,0],[3,1,0],[2,0,1],[1,2,0],[0,1,1] \, ,
\end{eqnarray}
whose $G_2$ Weyl dimensions are respectively, 1, 7, 14, 27, 64.

The relevant NIM-rep graph is illustrated in Figure
\ref{fig1}. The charge equations (\ref{twisteqn}) with $\lambda=\Lambda _1$ (fundamental representation of $D_4$ with dimension 8) thus are
\begin{eqnarray}
8\mathrm{q}_0&=&\mathrm{q}_{0}+\mathrm {q}_1 \nonumber \\
8\mathrm{q}_1&=&\mathrm{q}_{0}+2\mathrm {q}_1+\mathrm{q}_{2} + \mathrm{q}_{3}\nonumber \\
8\mathrm{q}_2&=&\mathrm{q}_{1}+\mathrm {q}_2+\mathrm{q}_{3} + \mathrm{q}_{4}\\
8\mathrm{q}_3&=&\mathrm{q}_{1}+\mathrm {q}_2+2\mathrm{q}_{3} + \mathrm{q}_{4}\nonumber \\
8\mathrm{q}_4&=&\mathrm{q}_{2}+\mathrm{q}_{3} + \mathrm{q}_{4}\nonumber
\end{eqnarray}

The first three equations are identically true with $\mathrm{q}_0=1,\mathrm{q}_1=7,\mathrm{q}_2=14,\mathrm{q}_3=27$, and $\mathrm{q}_4=64$, and the last two equations are satisfied
modulo $M_{D_4}=11$.

\end{document}